\documentclass[twocolumn,secnumarabic,amssymb, nobibnotes, aps, prc,superscriptaddress]{revtex4-1}

\usepackage{graphicx}
\usepackage{longtable}
\begin{document}
\title{\it{\textbf{Reaction and fusion cross sections for the near-symmetric 
system $^{129}Xe+^{nat}Sn$ from $8$ to $35$ $AMeV$.}}}

\author{L. Manduci}
\affiliation{Ecole des Applications Militaires de l'Energie Atomique, BP 19 50115, Cherbourg Arm\'ees, France}
\affiliation{Laboratoire de Physique Corpusculaire, ENSICAEN, Universit\'e de Caen, CNRS/IN2P3, F-14050 Caen Cedex, France}
\author{O. Lopez}
\affiliation{Laboratoire de Physique Corpusculaire, ENSICAEN, Universit\'e de Caen, CNRS/IN2P3, F-14050 Caen Cedex, France}
\author{A. Chbihi}
\affiliation{Grand Acc\'el\'erateur National d'Ions Lourds, CEA and CNRS/IN2P3, B.P.~5027, F-14076 Caen Cedex, France}
\author{M.F. Rivet}
\thanks{Deceased}
\affiliation{Institut de Physique Nucl\'eaire,CNRS/IN2P3, Universit\'e Paris-Sud 11, F-91406 Orsay Cedex, France}
\author{R. Bougault}
\affiliation{Laboratoire de Physique Corpusculaire, ENSICAEN, Universit\'e de Caen, CNRS/IN2P3, F-14050 Caen Cedex, France}
\author{J.D. Frankland}
\affiliation{Grand Acc\'el\'erateur National d'Ions Lourds, CEA and CNRS/IN2P3, B.P.~5027, F-14076 Caen Cedex, France}
\author{B. Borderie}
\affiliation{Institut de Physique Nucl\'eaire,CNRS/IN2P3, Universit\'e Paris-Sud 11, F-91406 Orsay Cedex, France}
\author{E. Galichet}
\affiliation{Institut de Physique Nucl\'eaire,CNRS/IN2P3, Universit\'e Paris-Sud 11, F-91406 Orsay Cedex, France}
\affiliation{Conservatoire National des Arts et Metiers, F-75141 Paris Cedex 03, France}
\author{M. La Commara}
\affiliation{Dipartimento di Scienze Fisiche and Sezione INFN, Universit\'a di Napoli "Federico II", I-80126 Napoli, Italy}
\author{N. Le Neindre}
\affiliation{Laboratoire de Physique Corpusculaire, ENSICAEN, Universit\'e de Caen, CNRS/IN2P3, F-14050 Caen Cedex, France}
\author{I. Lombardo}
\affiliation{Dipartimento di Scienze Fisiche and Sezione INFN, Universit\'a di Napoli "Federico II", I-80126 Napoli, Italy}
\author{M. P\^arlog}
\affiliation{Laboratoire de Physique Corpusculaire, ENSICAEN, Universit\'e de Caen, CNRS/IN2P3, F-14050 Caen Cedex, France}
\author{E. Rosato}
\thanks{Deceased}
\affiliation{Dipartimento di Scienze Fisiche and Sezione INFN, Universit\'a di Napoli "Federico II", I-80126 Napoli, Italy}
\author{R. Roy}
\affiliation{Laboratoire de Physique Nucl\'eaire, Universit\'e Laval, Qu\'ebec, Canada G1K 7P4}
\author{G. Verde}
\affiliation{Institut de Physique Nucl\'eaire,CNRS/IN2P3, Universit\'e Paris-Sud 11, F-91406 Orsay Cedex, France}
\affiliation{INFN - Sezione Catania, via Santa Sofia, 64, 95123 Catania, Italy}
\author{E. Vient}
\affiliation{Laboratoire de Physique Corpusculaire, ENSICAEN, Universit\'e de Caen, CNRS/IN2P3, F-14050 Caen Cedex, France}
\vspace{0.5cm}
\collaboration{INDRA collaboration}
\begin{abstract}
\begin{description}

\item[Background]Heavy-ion reactions from barrier up to Fermi energy.
\item[Purpose]Reaction and fusion cross sections determination.
Fusion reactions induced by $^{129}Xe$ projectiles on $^{nat}Sn$ targets for energies ranging from
$8$ A.MeV to $35$ A.MeV were measured with the INDRA $4\pi$-array.\\
The evaluation of the fusion/incomplete fusion cross sections for the incident energies from 8 to 35 A.MeV is the main purpose of
this paper.
\item[Method]
The reaction cross sections are evaluated for each beam energy thanks to INDRA $4\pi$-array.
The events are also sorted in order to focus the study on a selected
sample of events, in such a way that the fusion/fusion incomplete cross section is estimated. 
\item[Results] The excitation function of reaction and fusion cross sections were measured for the heavy and nearly symmetric system 
$^{129}Xe + ^{nat}Sn$ from 8 to 35 A.MeV.  
\item[Conclusions] The fusion-like cross-sections evaluated show a good agrement with a recent systematics for beam energies greater than
20 A.MeV. For low beam energies the cross-section values are lower than the expected ones. A probable reason for these low values is in
the fusion hindrance at energies above/close the barrier. 
\end{description}  
\end{abstract}
%

\maketitle
\section{Introduction}

Collisions between heavy ions at low energy above the barrier are dominated by binary inelastic collisions \cite{udo, glassel,charity,stefanini,wilczynski}. 
According to the prediction of the classical potential model of Bass, applied to the fusion of heavy nuclei, the limiting value 
for fusion is given for projectile and target combinations whose product $Z_pZ_t$ is not too large ($Z_pZ_t \leq 2500 - 3000$) 
\cite{bass0,NGO}. In this case the attractive pocket in the internuclear potential still prevents, for angular momentums $l \leq l_{crit}$, 
the reseparation of the dinuclear system, allowing it to evolve towards a compact shape and fusion occurs leading to compound nuclei 
with $Z \leq Z_p + Z_t$. The experimental signature of fusion processes consists in the presence of evaporation
residues and fission fragments in the exit channel.
For increasing projectile mass, the critical angular momentum increases. It can reach values larger than the one at which 
the fission barrier of the compound
nucleus vanishes \cite{bass0, tam}. As a consequence, 
the fusion cross 
section is expected
to fall to a negligible fraction \cite{mosel,lefort} since the Coulomb repulsion dominates and the potential well is not 
anymore able to trap the colliding nuclei and lead the system towards fusion. \par    
Therefore, one expects, as main exit channels, very dissipatives collisions \cite{udoB, wilcz, wilczb} and, with a reduced 
probability, fusion followed by two fragments emission. In the DIC the projectile and the target are strongly slowed down.
During the formation of a dinuclear system and before the reseparation, nucleons may be exchanged. The process lifetime (shorter than the
compound nucleus formation one) is deduced from the rotation
angle of this system before decay and the dissipated energy is function of the rotation angle \cite{toke}.\par 
Experimentally it was observed 
that the fission cross section was greater than the upper bound imposed by the presence of the minimum in the ion-ion potential \cite{heush}. 
Moreover the fission mass
distributions were wider than what expected on the basis of the compound nucleus model \cite{bernard, back}. Therefore, 
part of the cross section was ascribed to fusionlike processes characterized as "fission without barrier" which 
did not proceed through a compound formation. They are now referred to as quasi-fission \cite{toke, pete, itkisa, itkis}. Their 
interaction time is longer than the DIC phenomenon \cite{zag}. These capture reactions are practically indistinguishable from true compound
fission and without the knowledge of the interaction times it is not possible to establish whether the two detected fragments 
were generated in a true fission process or in a fast-fission one.
Moreover studies on the fusion-evaporation cross section for near symmetric systems \cite{Sahm, keller, morawek, Quint} 
provided evidence for the 
dynamical suppression of complete fusion even if the suppression may in some cases be due to a reduced detection efficiency. The concept of an "extra-push" in the interaction, conceived by Swiatecki, 
\cite{swia1, swia2, bloki} was necessary to allow the achievement of complete fusion. It was 
introduced in the interaction in form of one-body dissipation and experimentally consists in a shift of the effective mean fusion
barrier causing fusion hindrance in heavy systems at energies around the barrier. 
Dynamical fusion theories based on different approaches \cite{luk,ber} were able to reproduce data for fusion cross sections for
reactions between nuclei nearly symmetric with medium heavy masses ($A\simeq 100$). 
A recent model \cite{Swia2003}
gives the probability of compound nucleus formation as composed by the probability of formation in overcoming the ion-ion barrier all together to
the probability of diffusion toward a spherical shape from the dinuclear initial stage.     \par 

For incident energies at 10 or more  MeV/nucleon above the barrier, the appearance of pre-equilibrium nucleon emission gives rise to 
incomplete fusion processes leading to formation of compound nuclei with $A < A_p + A_t$. Moreover, since for higher beam energies 
more energy is converted into excitation
energy, events with three or four fragments in the exit channel may constitue an important fraction of the
associated cross section. 
\par  
INDRA \cite{pouthas1,pouthas2,parlog1,parlog2} has been used to perform a large body of 
measurements of the $^{129}Xe+^{nat}Sn$ system over a wide range of energies. This gave an unique opportunity for an important and 
exclusive study of reaction mechanisms for such a heavy quasi-symmetric system. \\
Previous works on data acquired with INDRA and concerning the same system at around and above Fermi energies 
\cite{lukasik,nat1,nat2,Neb,gou,ger,boc,meti,bor,john1,steck,belle,cussol,hudan1,colin,fev,pia,john2,pinch,lau,tab,nicol2007,pia1,fev1,
bonnet,lopezo,bonnet1,bor1,gru,lop2014,ade,gru1} focused mostly on the multifragmentation of a composite system formed in central collisions. 
In this
paper we study the energy range from just above the barrier ($8$ A.MeV) to the Fermi energy domain ($35$ A.MeV). First,
total reaction cross-sections are determined as a function of incident energy and compared with existing systematics. Then we present a new
method to estimate the total cross-section for both complete or incomplete fusion and capture reactions leading to fast or quasifission.
The resulting excitation function is compared with the recent systematics of \cite{nuovoeudes}.


\section{The Experiment}

The present study concerns the analysis of the data recorded during the $5^{th}$ INDRA 
campaign for reactions induced by $^{129}Xe$ projectiles on self-supporting 
350 $\mu g/ cm^2$
thick $^{nat}Sn$ targets at different beam energies E/A = 8, 12, 15, 18, 20, 25, 27, 29 and 
35 A.MeV.\par 
The experiment was performed at Ganil facility (Caen, France). Since the coupling of two main
cyclotrons (CSS1 and CSS2) did not allow to explore the whole incident energy range, the
$^{129}Xe$ beam was first accelerated at 27 A.MeV and successively degraded, through carbon foils
of different thickness, to the energies of interest. The charge state of the primary beam was
40+. After the degrader, as expected, the Xe beam had a wide distribution of charge states.
Therefore, with the help of the $\alpha$-spectrometer, only one charge state was selected. The 
B$\rho$ setting of the spectrometer was optimized for each incident energy. However at low energy more than 
one charge state was transmitted and this affected the incident energy
with uncertainties around $\Delta E \simeq 1$ MeV for the beam energies at 8 and 12 A.MeV. The energies at E/A = 29 and 35 A.MeV were obtained by direct tuning. \\
INDRA is a charged particle multidetector covering $90\%$ of the total solid angle.
It is composed by 336 independent telescopes arranged in 17 rings centered on the beam axis. 
In the first ring ($2^{\circ}$ and $3^{\circ}$) are arranged 12 telescopes composed of a 300 $\mu$m
silicon wafer and a CsI(Tl) scintillator crystal (14 cm thick). Rings 2 to 9 
($3^{\circ}$ to $45^{\circ}$) have 12 or 24 three-stage detection telescopes : a
gas-ionization chamber (filled with $C_3F_8$), a 300 or a 150 $\mu$m silicon wafer and a CsI(Tl) scintillator 
(14 to 10 cm thick) coupled to a photomultiplier tube. Rings 10 to 17  
($45^{\circ}$ to $176^{\circ}$) are comprised of 24, 16 or 8 two-member telescopes : a
gas-ionization chamber and a CsI(Tl) scintillator of 8, 6 or 5 cm thick. A more detailed description may
be found in references \cite{pouthas1, pouthas2, parlog1, parlog2}.\\
\begingroup
\squeezetable
\begin{table}
\begin{center}
\begin{tabular}{|c|c|c|c|c|c|}
\hline
$E_{inc}$ $(A.MeV)$ & $E_{CM}$ $(MeV)$ & $E_{CM}/V_{C}$&$v_{Lab}$ &
$v_{CM}$ & $\Theta_{gr}^{\circ}$   \\ 
\hline
$8$  & $494.6$&1.8&$3.90$&$2.04$&$22.13$   \\
\hline
$12$  &$741.6$&2.7&$4.77$&$2.50$&$12.84$   \\
\hline
$15$  &$926.6$&3.4&$5.32$&$2.79$&$9.79$    \\
\hline
$18$  &$1111.5$&4.0&$5.81$&$3.05$&$7.91$    \\
\hline
$20$  &$1234.6$&4.5&$6.12$&$3.21$&$7.02$   \\
\hline
$25$  &$1542.3$&5.6&$6.81$&$3.59$&$5.47$    \\
\hline
$27$  &$1665.2$&6.0&$7.07$&$3.73$&$5.03$    \\
\hline
$29$  &$1788.1$&6.5&$7.31$&$3.86$&$4.65$   \\
\hline
$35$  &$2156.3$&7.8&$8.00$&$4.23$&$3.80$   \\
\hline
\end{tabular}
\caption{\label{cin} Kinematic characteristics for the $^{129}Xe+^{nat}Sn$ system at 
different incident energies. The laboratory velocity $v_{Lab}$ and the center mass velocity $v_{CM}$ are
in $(cm/ns)$.}
\end{center}
\end{table}
\endgroup
INDRA can measure ion charge and energy in a wide range and can resolve masses up to 
$Z=4$. The charge identification was realized by means 
of the $\Delta E$-$E$ matrices which well reproduces the
form of the lines for each atomic number, $Z$. Unit charge resolution
was obtained for all nuclei produced in this reaction.
The energy identification threshold is $\simeq 0.8 - 1$ MeV/nucleon for light fragments and around $\simeq 1.5-1.7$ MeV/nucleon for
fragments of $Z=50$. 
\\
Collision data for the $^{129}Xe+^{nat}Sn$ system at the various beam energies were recorded with an 
acquisition trigger requiring 1, 2, 3 or 4 fired telescopes in coincidence.\\
Table \ref{cin} shows the reaction kinematic characteristics for all beam energies. 
The Coulomb barrier for this system at interaction radius amounts to $V_{Coul} \simeq 275$ 
MeV. As it appears from the 
ratio of the available energy in the centre of mass $E_{CM}$ to the Coulomb barrier $E_{CM}/V_{Coul}$, 
in the third column of table \ref{cin}, all the reactions take place well above the  
barrier.\\

\section{Reaction Cross Section}

The total reaction cross section may be defined as the total cross section 
minus the elastic scattering contribution :
\begin{equation}
\sigma_R = \sigma_T - \sigma_{el}\label{crossi}
\end{equation}
In order to deduce the experimental reaction cross sections, data with trigger multiplicity 
$MULT \geq 1$ were analyzed and, under 
appropriate constraints, the elastic peak was isolated for each beam energy in order to evaluate 
 the elastic scattering cross section to be subtracted in equation (\ref{crossi}) from the total cross section 
 (see $\Theta_{gr}^{\circ}$ in table
 \ref{cin}). 
 This latter was computed as it follows :
\begin{equation}
\sigma_T = \frac{N_{event}}{N_tI}
\end{equation}
where $N_{event}$ is the total number of recorded events, $N_t$ the nuclear density of the 
target and $I$ the incident flux, particles per unit time, expressed as :
\begin{equation}
I = {\it{F}}\frac{(1-\tau)}{qe} 
\end{equation}
Here {\it{F}} is the charge measured by the Faraday cup, $\tau$ is the acquisition dead 
time expressed as a fraction of the total acquisition time, $q$ is the equilibrium value of the projectile charges
evaluated using the reference \cite{qu} and $e$ is the
elementary charge.\par
\begin{figure}[!h]
\begin{center}
\includegraphics*[scale=0.46]{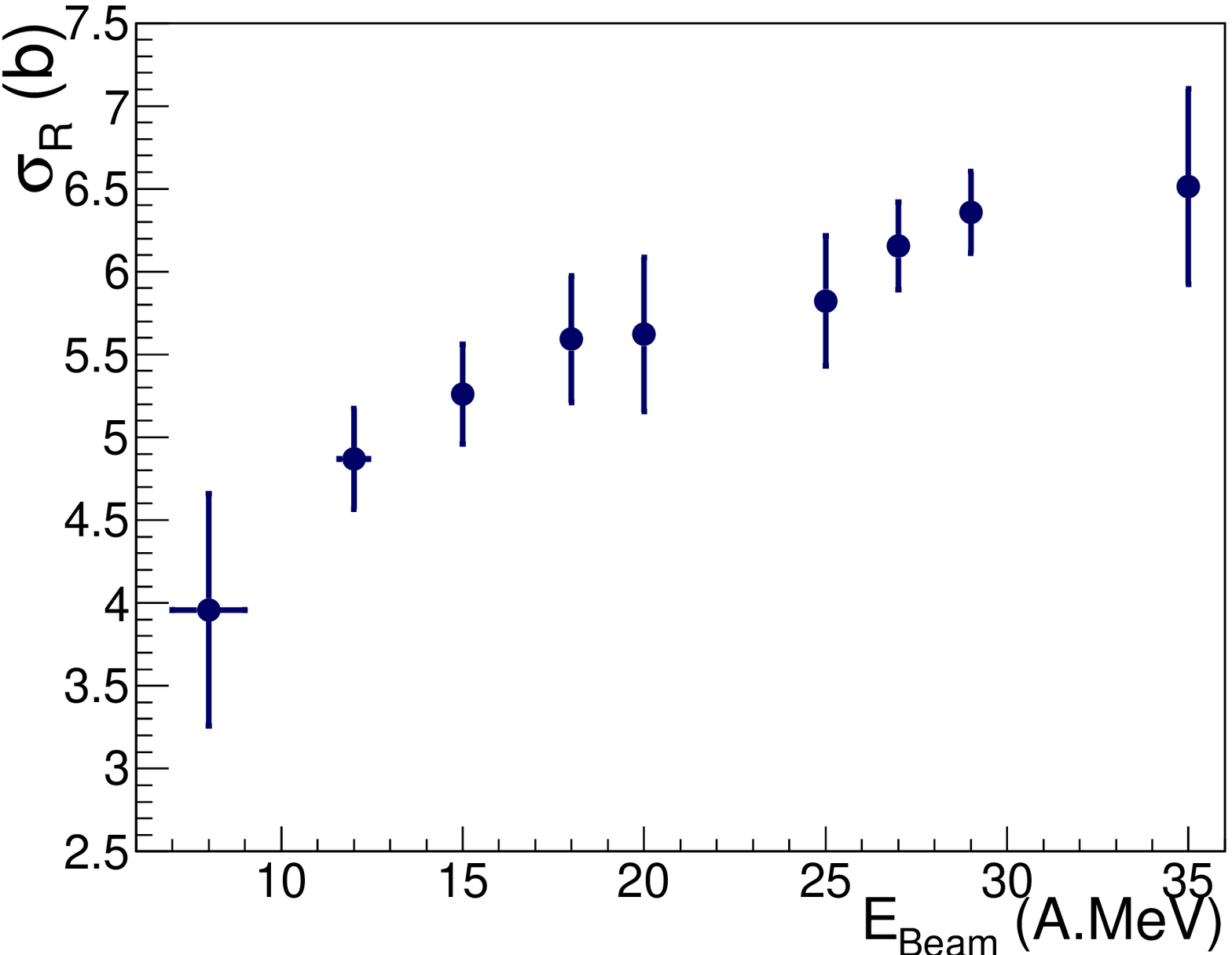}
\includegraphics*[scale=0.46]{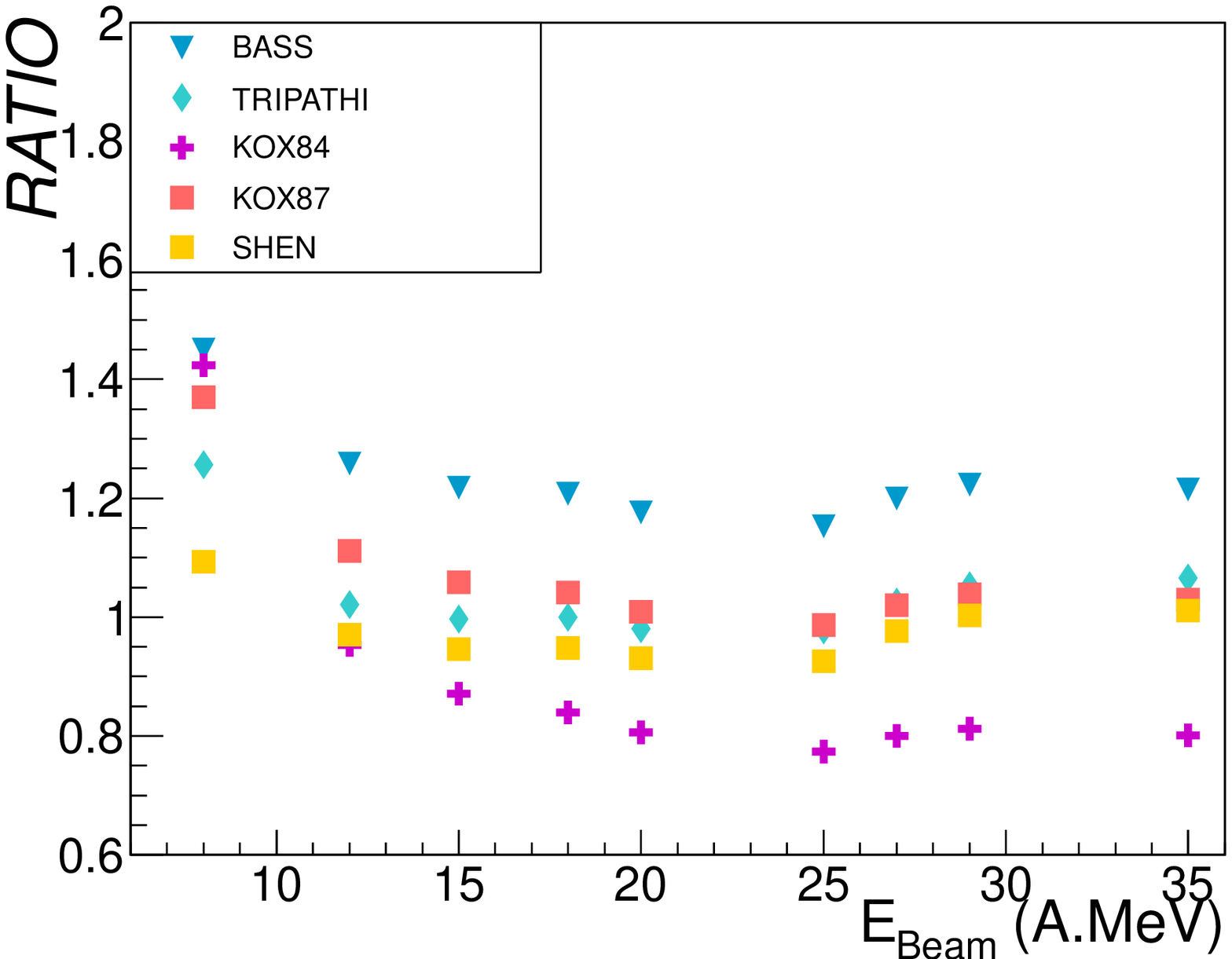}
\caption{Upper panel : experimental reaction cross sections for all beam energies. Lower panel : experimental reaction cross sections 
normalized to the theoretical values of the following parameterizations : Bass \cite{Bass}, Kox \cite{kox}, Tripathi \cite{TRI} 
and Shen \cite{shen}.}
\label{rcross}
\end{center}
\end{figure}

 
The experimental reaction cross section values obtained with this procedure are reported in table \ref{crossth} 
and displayed in the upper 
panel of figure \ref{rcross}. As one can see, they show a rapid increase with beam energy up to 20 MeV/nucleon and then tend
 toward an asymptotic limit close to of a purely geometrical cross-section.
\begin{table}[!h]
\caption{\label{crossth}Experimental and theoretical reaction cross sections in barns for each beam energy in
A.MeV.}
\begin{center}
\begin{tabular}{|c|c|c|c|c|c|}
\hline
 $E/A$& $\sigma_{r}^{Exp}$ &  $\sigma_{r}^{Bass}$ &$\sigma_{r}^{Kox87}$ &$\sigma_{r}^{Tripathi}$
 &$\sigma_{r}^{Shen}$  \\
\hline
 8&$3.96\pm 0.70$ &2.73&2.89&3.15&3.62\\
\hline
 12&$4.87\pm 0.30$ &3.87&4.38&4.77&5.02\\
\hline
15&$5.26\pm 0.30$&4.32&4.97&5.28&5.56\\
\hline
18&$5.59\pm 0.38$&4.63&5.37&5.60&5.90 \\
\hline
20&$5.62\pm 0.47$&4.78 &5.57&5.73&6.04\\
\hline
25&$5.82\pm0.39$&5.05&5.92&5.95&6.29 \\
\hline
27&$6.15\pm0.26$&5.13&6.03&6.00&6.30 \\
\hline
29&$6.36\pm 0.24$&5.20&6.12&6.04&6.34 \\
\hline
32&$5.46\pm 0.10$&5.28&6.24& 6.08&6.40\\
\hline
35&$6.51\pm0.59$&5.36&6.33&6.11&6.44 \\
\hline
\end{tabular}
\end{center}
\end{table}
The associated error bars are mainly due to the uncertainties
of the charge state for each incident energy. They reflect also the difficulty in some case to
accomplish a proper definition of the elastic peak. In particular, for the beam energy 35 A.MeV the error bars 
are larger because the elastic peak was mostly lost as consequence of the small grazing angle 
(see $\Theta_{gr}^{\circ}$ in table \ref{cin}). The error bars associated to the beam energy are also shown for 8
and 12 A.MeV.\\
In the lower panel of figure \ref{rcross} the experimental reaction cross sections were normalized to the ones obtained by 
different theoretical parameterizations from Bass \cite{Bass}, Kox \cite{kox}, Tripathi \cite{TRI} and Shen \cite{shen} 
reported in table \ref{crossth}.
The Bass parameterization was deduced in the classical framework of the strong absorption model and does not contain any mechanism of
energy dissipation. As can be seen from the figure, this parameterization can constitute a lower bound for the reaction cross section of
our system while the one labelled Kox84 can be considered as an upper bound (see also table \ref{crossth}).
The best agreement with data is found for those parameterization (labelled in figure as Kox87, Tripathi and Shen) in which were 
introduced  corrections for the 
neutron excess skin Kox et al. \cite{kox}, for the transparency
and the Pauli blocking Tripathi et al. \cite{TRI}. The parameterization in Shen et al. \cite{shen} uses an unified formula 
from low to intermediate energies. It
will be used when computing the fusion cross section for normalization as in the reference \cite{nuovoeudes} 
for comparison.\\ 

\section{Fusion and Incomplete Fusion cross section evaluation}  

In this section we will evaluate the fusion cross-section for the collision system at all incident energies. We discuss in a first time the global
observable $E_{iso,max}$ and the selected data with the suitable characteristics for the cross section evaluation. A comparison with
recent analysis in reference \cite{nuovoeudes} will also be exposed.  

\subsection{Observable $E_{iso,max}$}

In order to select classes of events with marked fusion characteristics, 
\begin{figure}[!h] 
\begin{center}
\includegraphics*[scale=0.44]{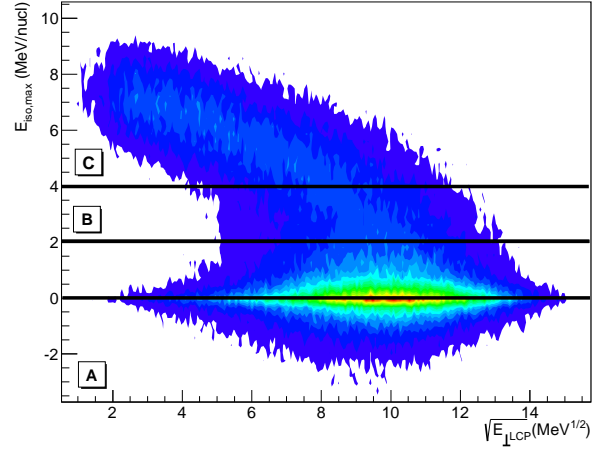}
\caption{$E_{iso,max}$ versus $\sqrt{E_{\perp LCP}}$ for the system at 15 A.MeV. $E_{iso,max}$ appears subdivides in three zones :
A, B and C. See the text to understand their meaning.}
\label{ORDIASCI}
\end{center}
\end{figure}
\begin{figure}[!h] 
\begin{center}
\includegraphics*[scale=0.44]{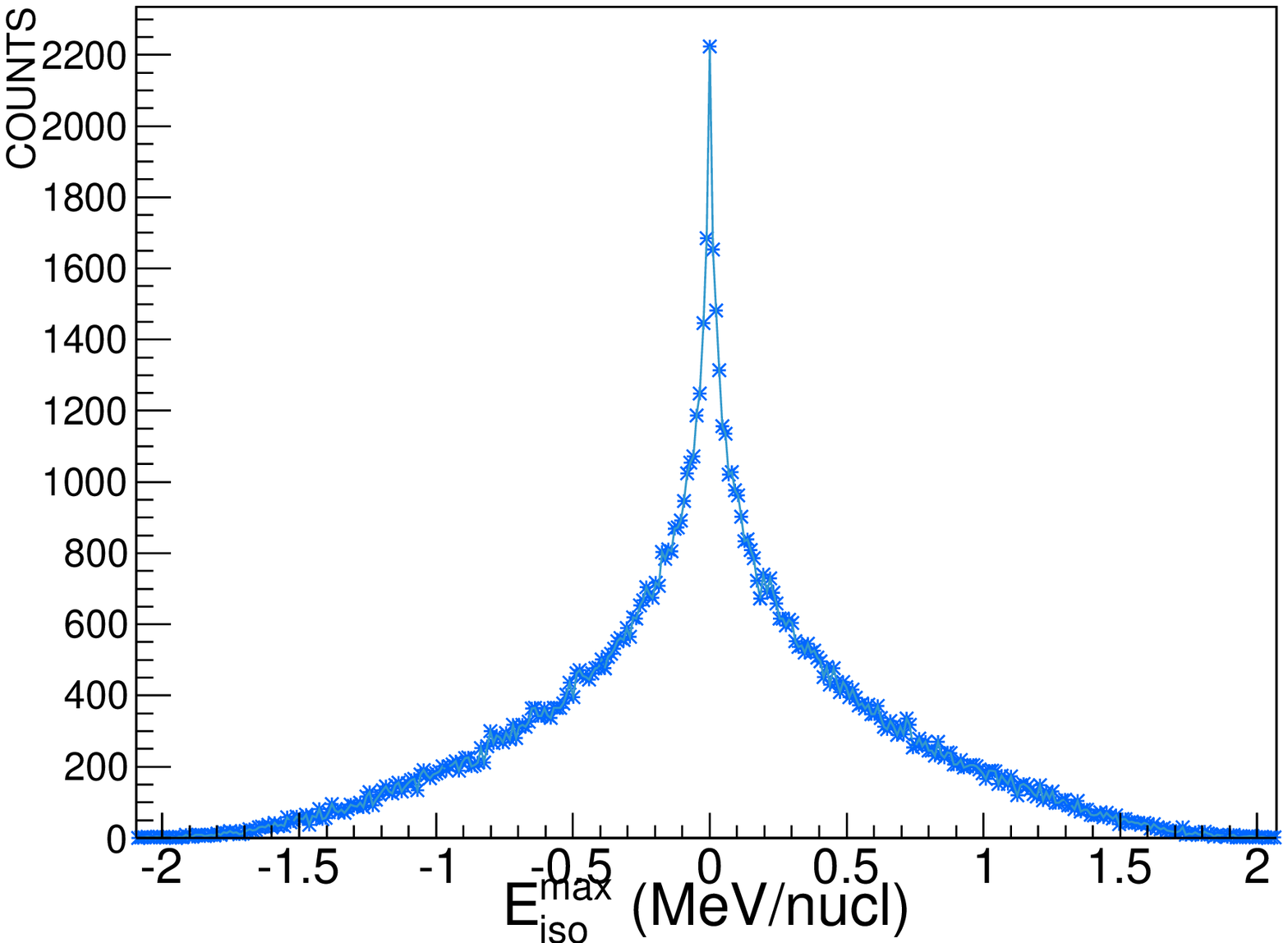}
\includegraphics*[scale=0.44]{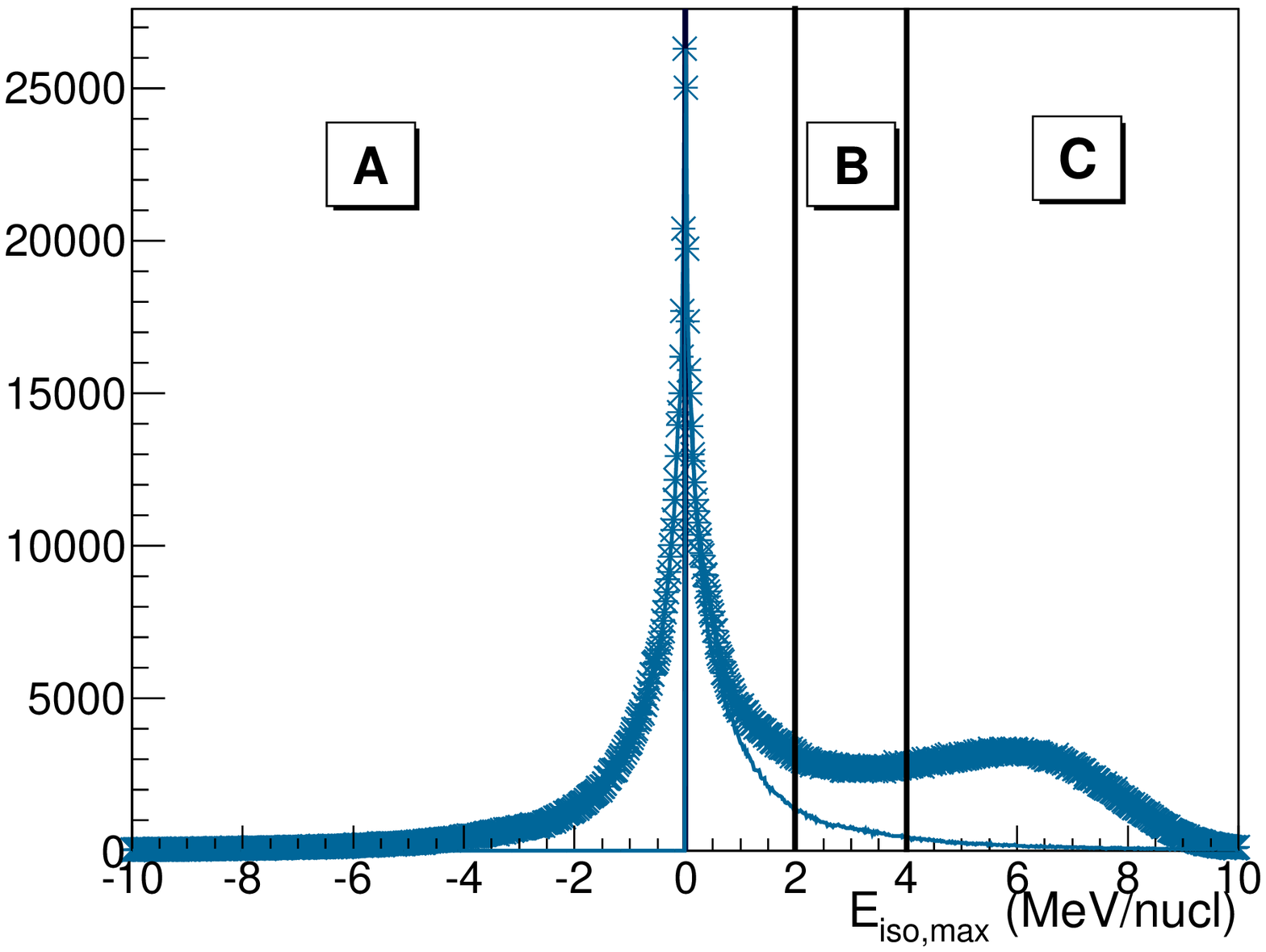}
\caption{Upper panel : simulation of $E_{iso,max}$ for pure central collision with unique source formation. In this case the observable 
is perfectly symmetric around zero \cite{remi, john}. \\Lower panel :
experimental $E_{iso,max}$ at 15 A.MeV.  In order to appreciate this symmetry, more hidden for experimental data, 
the negative part has been reversed and superimposed to the positive values. The labels A,B and C refer to the zones commented in
the text.}
\label{eisoSIMMretta}
\end{center}
\end{figure}
%
%
%
 the kinematic global observable $E_{iso,max}$ \cite{pippo,remi} in term of the velocity components of the 
heaviest fragment in the event was used. This observable is defined as :
\begin{equation}
E_{iso,max} = V_{\parallel,max}^{2}-0.5V_{\perp,max}^{2}(1+sin2\phi)\label{ORDI}
\end{equation}
where $V_{\parallel,max}$ and $V_{\perp,max}$ are the velocity components of the heaviest fragment in the centre of mass (CM) parallel and
orthogonal to the beam direction and $\phi$ is its azimuthal
angle.\\ 
The $E_{iso,max}$ observable enhances the
separation between the projectile-like contributions and the more damped events whose products are produced at
rest in the centre of mass frame. The upper panel of figure \ref{eisoSIMMretta} shows the result of a simulation in which fragments have
an isotropic momentum distribution in the centre of mass frame \cite{john}. The resulting distribution of $E_ {iso}^{max}$ is symmetric
around zero, even for events with only two fragments (fission-like events). If the source of emission is not at rest in the centre of mass
frame, but moves with a moderate velocity ($\leq 1$ $cm/ns$), either faster or slower than the centre of mass frame (such as in the case
of incomplete fusion), the distribution is still peaked at zero but skewed towards positive values of $E_ {iso}^{max}$, in such a way that
the total number of events with $E_ {iso}^{max} < 0$ is less than $50\%$ of the total. A similar effect is observed for a non-isotropic
emission pattern at rest in the centre of mass. On the other hand, for larger source velocities in the CM frame (such as for
projectile-like decays), the whole distribution is shifted to positive values without a pronounced peak and there are no longer any events
for which $E_ {iso}^{max} < 0$. Therefore the measured cross-section for $E_ {iso}^{max} < 0$ can be considered as a lower limit for the
cross-section for capture reactions (fusion-evaporation, fusion-fusion, quasifission), with a negligible contribution from binary
dissipative collisions. A simulation on the present collision system for central events with the code SMM \cite{bot} 
gave the same pattern \cite{remi} as the upper panel of the figure \ref{eisoSIMMretta}.\par 
In figure \ref{ORDIASCI} is shown the experimentally measured correlation between $E_{iso,max}$ and the quantity $\sqrt{E_{\perp LCP}}$
for 15 A.MeV bombarding energy. $\sqrt{E_{\perp LCP}}$ is the square root of the total transverse energy of light charged particles 
($Z < 3$) and is related to the degree of centrality of the collision (\cite{phair}, \cite{pla}). One can identify two components,
separated by the black line in the figure at $E_ {iso}^{max}\simeq 2$. The first (labelled with A) is the component with $E_{iso,max} < 2 MeV/A$. 
The second for values $E_{iso,max} > 2 MeV/A$, comprises the two zones labelled
with B and C. These two components 
indicate clearly an evolution of the dissipated energy from central to peripheral collisions. The
deep valley observed close to $E_ {iso}^{max}\simeq 2$ helps to accomplish the separation between binary
(deep inelastic collisions, DIC) and central collisions (candidate for fusion).\par
The lower panel of figure \ref{eisoSIMMretta} shows the experimental observable $E_ {iso}^{max}$ for
the collision at 15 A.MeV. This correlation is not symmetric around zero since it contains all the reaction
contributions. To guide the eye the negative part was reversed and superposed to the positive one. According to the results of the
simulation presented above, in the following we will estimate the fusion-like cross-section by doubling the yield of events 
with $E_ {iso}^{max} < 0$.

%

\subsection{Event selection by $E_{iso,max}$ bins.}
The global observable $E_{iso,max}$ may be used to sort the events accordingly to the underlying reaction 
mechanism. In fact, depending on the choice of the bins in which the observable can be 
divided, it is 
possible to select roughly three classes of events : one for which the fusion-like/capture reactions are the dominant 
mechanism; a second resembling highly-damped binary collisions; and finally events belonging to less 
dissipative reactions. \\
\begin{figure}[!h] 
\begin{center}
\includegraphics*[scale=0.45]{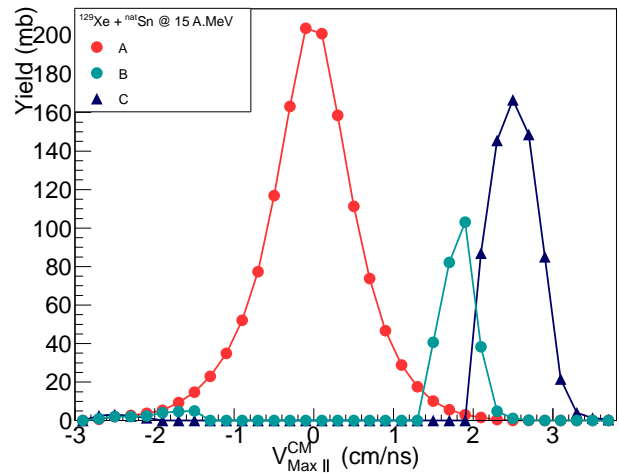}
\caption{Heaviest fragment parallel velocity for three different bins of $E_{iso,max}$ for 
the system at 15 A.MeV.  $A$ represents event selected with $E_{iso,max} \leq 0$, $B$ with 
$2 \leq E_{iso,max} \leq 4.$ and $C$ with $E_{iso,max}> 4.$}
\label{Vmaxordi}
\end{center}
\end{figure}
\begin{figure}[!h] 
\begin{center}
\includegraphics*[scale=0.44]{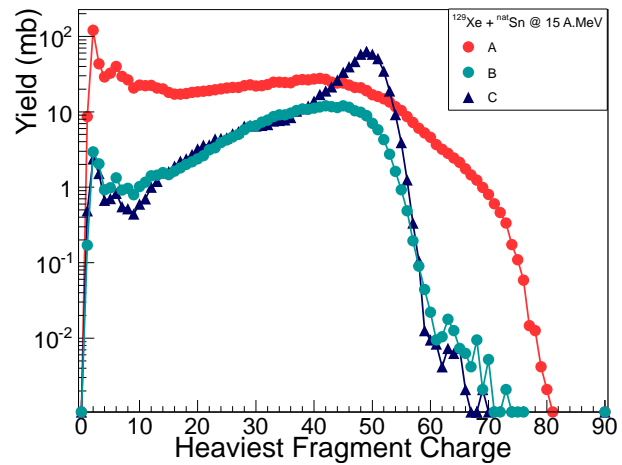}
\caption{Heaviest fragment charge for bins of $E_{iso,max}$ as figure \ref{Vmaxordi} for 15 A.MeV. The distributions are
expressed in mb. }
\label{zmaxordi}
\end{center}
\end{figure}
Figure \ref{Vmaxordi} shows the longitudinal velocity distributions 
of the heaviest fragment for events selected according to $E_{iso,max}$ bins values for 
the reaction at 15 A.MeV. The events labelled with $A$ were 
selected via $E_{iso,max} \leq 0$ and doubled. In this case the distribution is gaussian and symmetric around 
zero and groups fusion events. The distribution labelled with $B$ is formed by events 
selected with $2 \leq E_{iso,max} \leq 4.$ and is subdivided into two asymmetric bumps.  
These two bumps should actually have the same size but due to identification thresholds at backward angles for the slow-moving
quasi-target, the latter gives just a very small bump at negative centre of mass velocities.
The third component, for $E_{iso,max}>4.$  is close to the beam velocity, indicating  collisions with little dissipation, 
for which the Quasi-Projectile (QP) was detected.
 \par  
Figure \ref{zmaxordi} shows, for the same bins of $E_{iso,max}$ as in figure \ref{Vmaxordi}, the
charge distributions of the heaviest fragment expressed in millibarns. As before, selection A gives the charge
distribution for the heaviest fragment for the fusion-like events. From this distribution is clear that the condition 
$E_{iso,max} \leq 0$ selects also events without fragments (in this work are named fragments nuclei with 
charge greater than 10). This means that events just constituted by light charged particles and
intermediate mass fragments with charge $3 \leq Z \leq 9$ are also included. In these events the heaviest
fragment can be a proton, an alpha or an IMF. At low excitation energies these events are issued by evaporation from 
the composite system or by one of the two partners in a DIC event. The condition $E_{iso,max} \leq 0$ applied to these
events is able to select properly the fusion events. Moreover, till 18 A.MeV, the events selected either 
with zero fragment multiplicity or 
with both conditions : $E_{iso,max} \leq 0$ and zero fragment multiplicity do not differ too much, since the non-fusion
contribution is small.   
Starting from 20 A.MeV, a larger number of IMF is produced from neck fragmentation \cite{DiToro}. As a consequence, 
the events with no fragments show a velocity 
distribution more centered around very low velocities (one can speculate that these light charged particles are 
 mostly coming from the target evaporation). The selection $E_{iso,max} \leq 0$ still selects fusion events but 
 the velocity distributions are not anymore gaussian : they are slighty asymmetric toward lower
 velocities. This contribution may be removed asking a further constraint on the heaviest fragment charge, as it will
 be explained in the next paragraph.\\ 
\begin{figure}[!h] 
\begin{center}
\includegraphics*[scale=0.45]{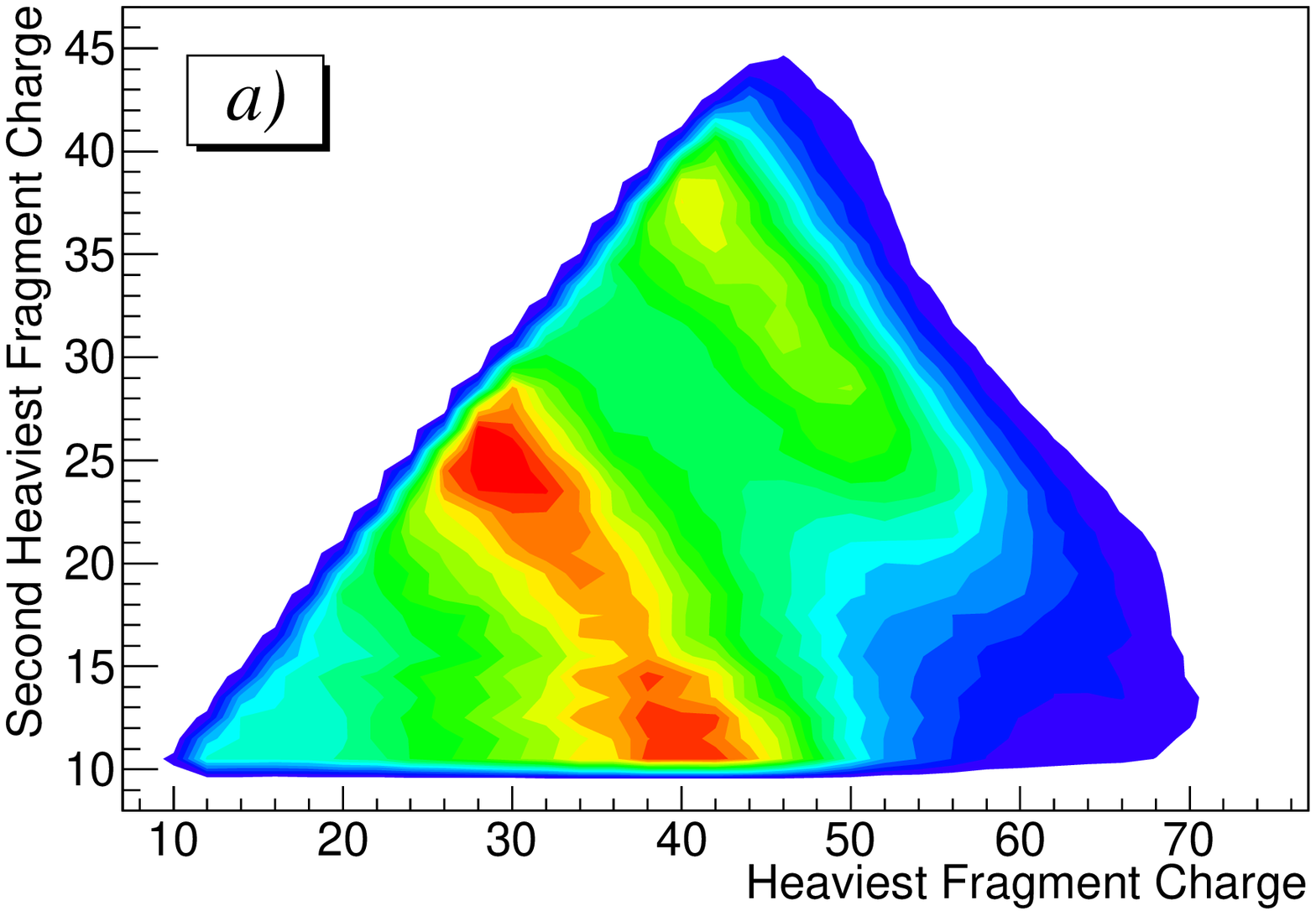}
\includegraphics*[scale=0.45]{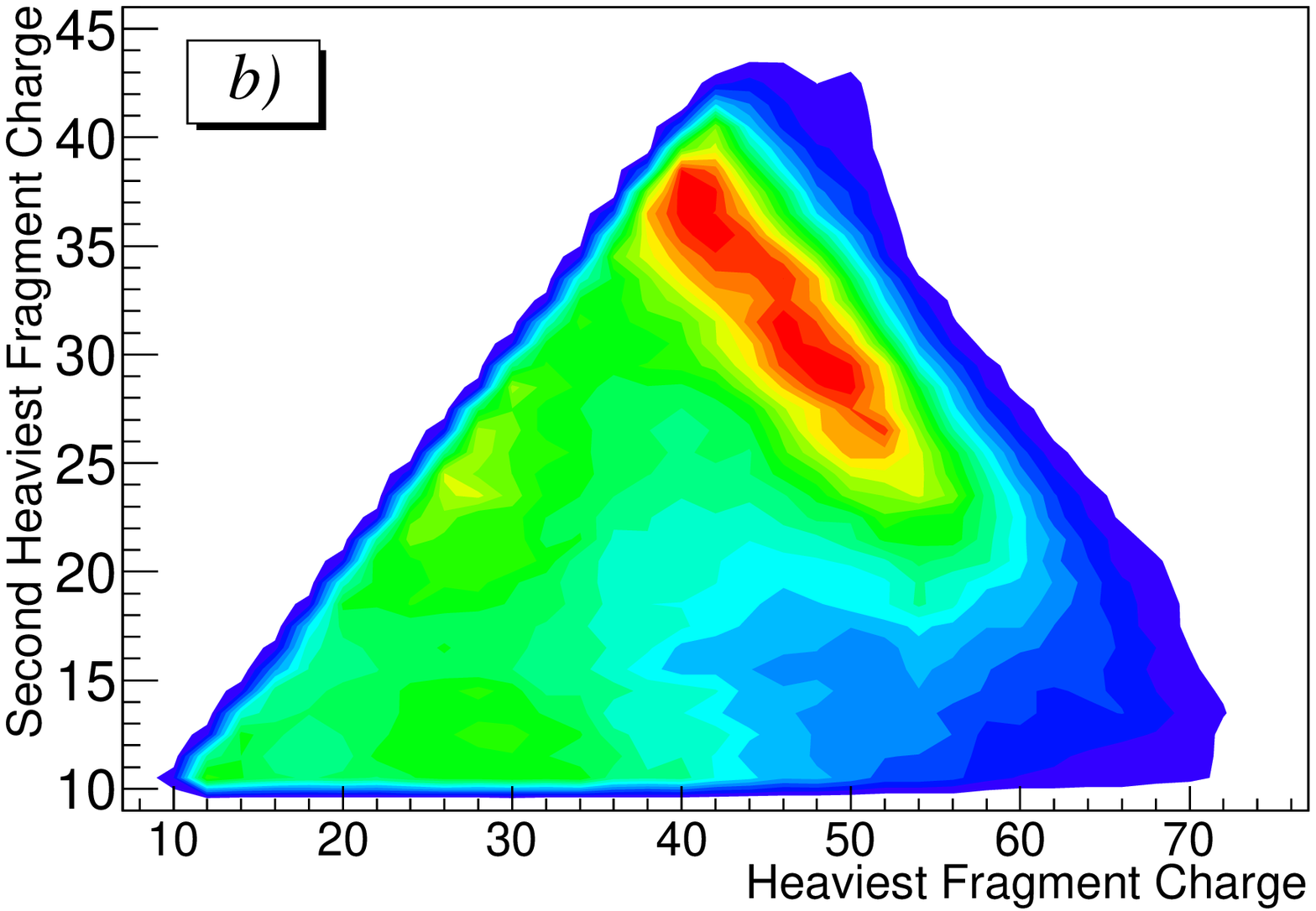}
\caption{Correlation of the second heaviest fragment charge with the heaviest fragment charge for the
system at 15 A.MeV for events with two fragments in the exit channel. $a)$ no selection on $E_{iso,max}$. $b)$ $E_{iso,max} \leq 0$.}
\label{Z1Z2}
\end{center}
\end{figure}
Figure \ref{Z1Z2} shows the correlation of the heaviest and second heaviest fragment charges for events with two 
fragments in the exit channel. In
the panel $a)$ the correlation is without selection on $E_{iso,max}$. The panel $b)$ shows the correlation for $E_{iso,max} \leq 0$. 
The constraint on the global variable selects therefore events in which the two fragments result from the scission of a composite system 
with $Z_1+Z_2 \simeq 75-80$. In fact the
lower ridge of figure $a)$ corresponds to events having a relative folding angle which is around $90-100^{\circ}$ while the higher ridge, 
which is more evident in the panel $b)$ after the selection with $E_{iso,max} \leq 0$, has a relative folding angle centered at 
$160^{\circ}$, close to back-to-back emission.\par
\begin{figure}[!h] 
\begin{center}
\includegraphics*[scale=0.45]{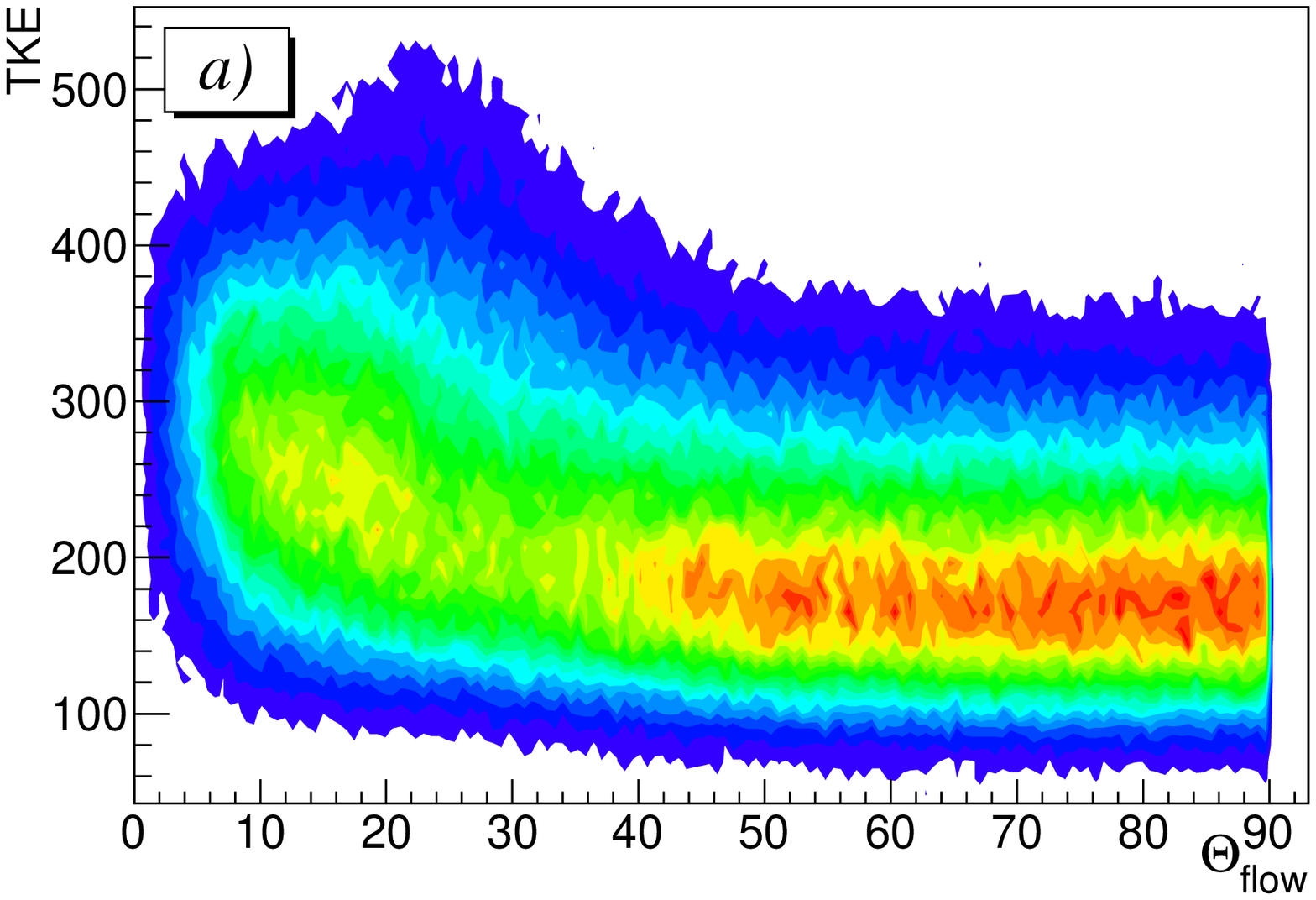}
\includegraphics*[scale=0.45]{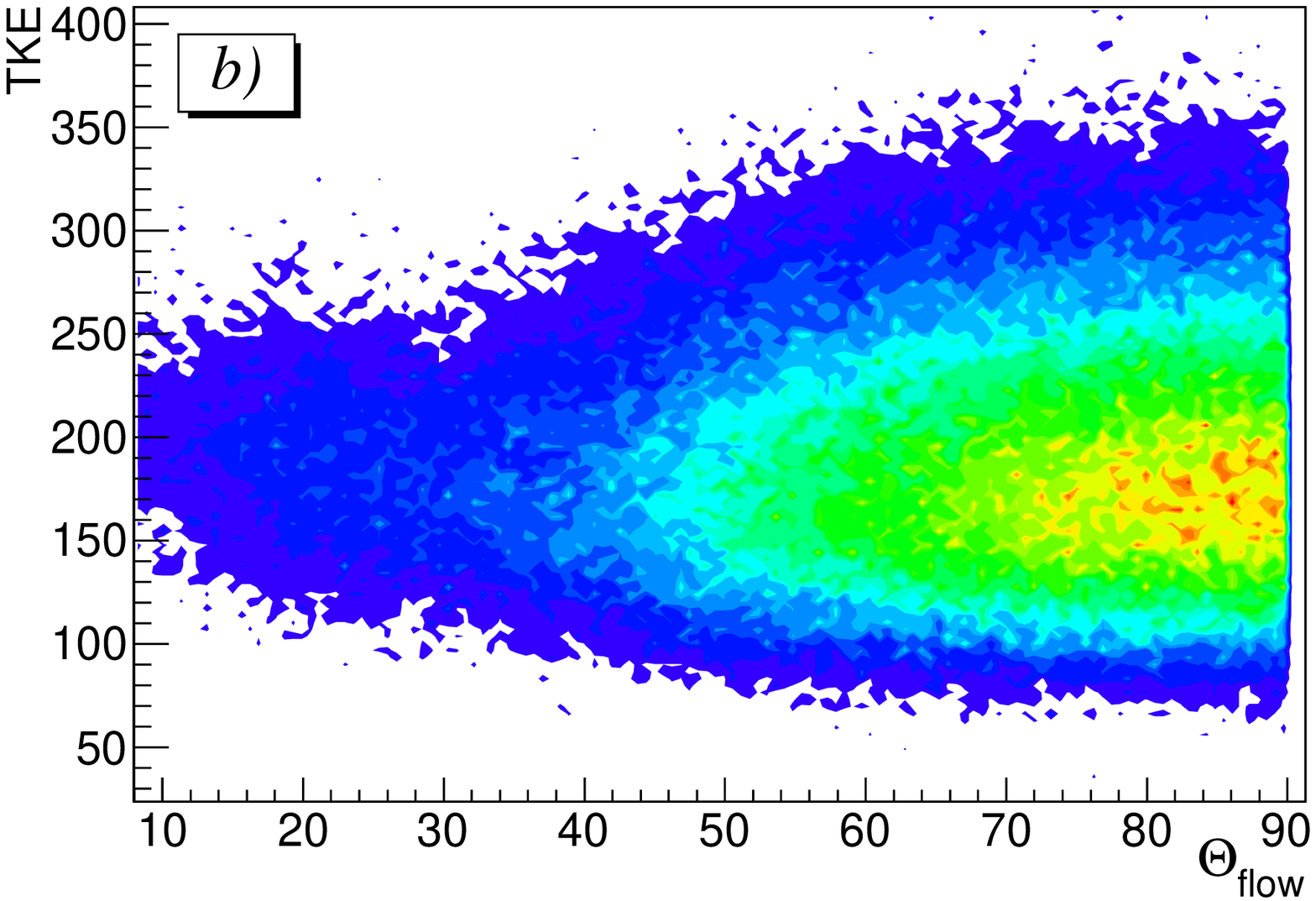}
\caption{Total kinetic energy versus flow angle for the
system at 15 A.MeV for events with two fragments in the exit channel. $a)$ no selection on $E_{iso,max}$. $b)$ $E_{iso,max} \leq 0$.}
\label{flusso}
\end{center}
\end{figure}
\begin{figure}[!h] 
\begin{center}
\includegraphics*[scale=0.44]{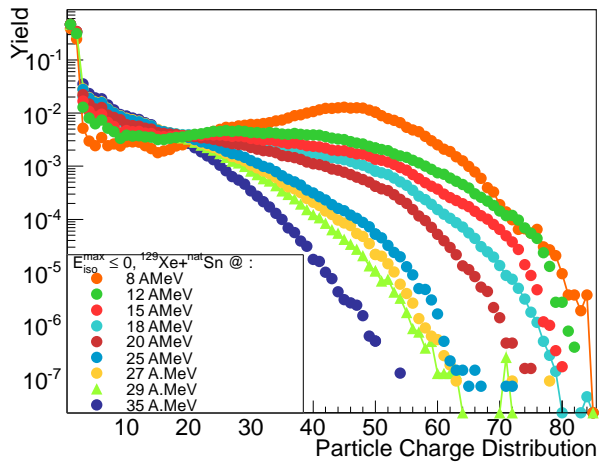}
\caption{Particle charge distributions at all beam energies. See text.}
\label{zi_fact}
\end{center}
\end{figure}

Figure \ref{flusso} shows the same selection applied to the Wylczinski plot : fragment total kinetic energy (TKE) in function of 
the flow angle \cite{wilcz,wilczb,flot}. The selection condition applied to data in the lower panel, figure $b)$ selects mainly events
 with a small TKE and a near-isotropic distribution of flow angles (peaked at $\Theta_{flow} \simeq 90^{\circ}$). It should be noted that
 the TKE distribution is peaked between the values expected for symmetric fission of composite systems with $Z_1+Z_2 \simeq 75-80$
 ($TKE_{SF} \simeq 140 MeV$) and $Z_1+Z_2 = 104$ ($TKE_{SF} \simeq 200 MeV$) \cite{viola}.\\  
We conclude this paragraph showing, in figure \ref{zi_fact}, the charge distributions for each incident energy. In this figure, 
the distributions were weighted in order to give the cross
section fraction pertinent to each beam energy and then normalized to the event number.
The $E^{max}_{iso} \leq 0$ selection was also applied. 
\subsection{Fusion cross-section evaluation.}

In this paragraph the attention will be focused on fusion/incomplete fusion reactions. We will evaluate the experimental
fusion cross sections by using the condition $E_{iso,max} \leq 0$ alone or adding one
more constraint and then doubled. 
The values of the fusion/incomplete fusion cross-sections
found in our analysis will be compared to the predictions of \cite{nuovoeudes} in which a function was
deduced from an experimental systematics based on the mass asymmetry parameter.\par
\begin{figure}[!h] 
\begin{center}
\includegraphics*[scale=0.44]{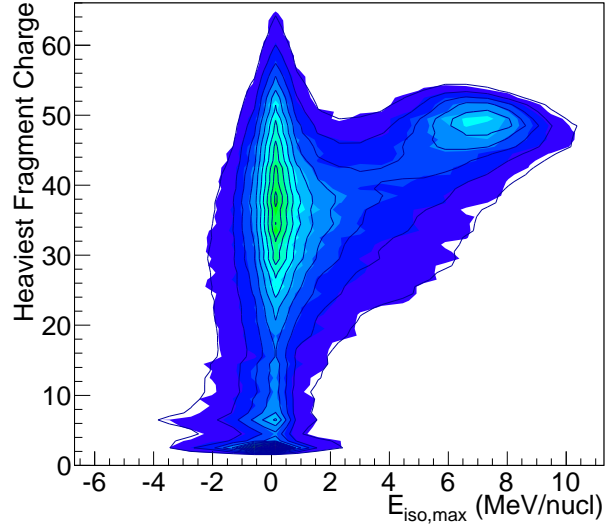}
\caption{Correlation for the heaviest fragment charge versus $E_{iso,max}$ for $E_{Beam}= 18$ A.MeV.}
\label{due}
\end{center}
\end{figure}
\begin{figure}[!h] 
\begin{center}
\includegraphics*[scale=0.44]{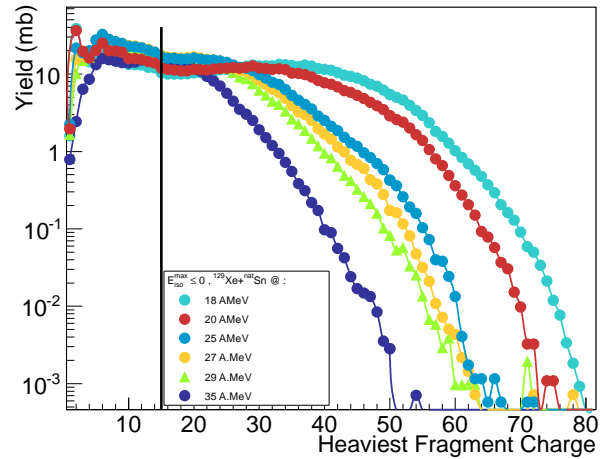}
\caption{Heaviest fragment charge at incident energies 18-35 A.MeV with the $E^{max}_{iso} \leq 0$ selection. The vertical black line gives the
cut at $Z_{max}=15$.}
\label{grossiCUT}
\end{center}
\end{figure}

To estimate the fusion/incomplete fusion cross-section, we also need to account for
those events for which not all the particles were completely detected. An initial selection based only on the computation of complete 
events would have
drastically excluded all the events where the residue or one of the fission fragments was lost. 
Consequently it was argued more correct to select fusion events by the condition $E^{max}_{iso} \leq 0$ for the reasons discussed above. 
These
events were then doubled according to the symmetry of the observable. We did however require that the heaviest fragment of each event was
identified in either the ionization chamber-silicon telescopes ($\theta \leq 45^{\circ}$) or in the ion chamber-CsI telescope 
($\theta > 45^{\circ}$), which excludes most events where the heaviest fragment is a light charged particle ($Z\leq 2$).\\
Starting from $E_{Beam}=18$ A.MeV up to $E_{Beam}=35$ A.MeV we added also a constraint on the heaviest fragment charge to reject events 
with increasing IMF multiplicities coming from the neck emission. Figures \ref{due} and \ref{grossiCUT} help to understand this point. 
Figure \ref{due}
shows the heaviest fragment charge versus the $E_{iso,max}$ at 18 A.MeV. In this figure one can observe three
bumps : one for peripheral events, one for fusion events and one mostly constituted by events in which the heaviest 
fragment has a very
low charge because the true one was not detected. Figure \ref{grossiCUT} shows the heaviest charge distributions for beam
energies from 18 to 35 A.MeV expressed in mbarns. The condition $E^{max}_{iso} \leq 0$ was applied. 
One can remark minima in the heaviest charge are evident, with a greater increase for higher energies. 
 On the basis of this feature  
it was considered as a better choice to accept 
events whose the
heaviest fragment charge was larger than a certain limit, deduced from figure  \ref{grossiCUT}. This limit, actually the same
for each beam energy, was set at $Z_{max} \geq 15$. In this way the contribution from the neck emission was minimized.
\\ 

%
\begin{table}[!h]
\caption{Fusion cross section values in $mb$.}
\label{fusion}
\begin{center}
\begin{tabular}{|c|c|}
\hline
$E_{beam}$ (A.MeV) & $\sigma_{Fus/IF} (mb)$  \\
 \hline
 8& $390\pm 50$\\
 \hline
 12&$752\pm130$\\
\hline
 15& $1100\pm100$\\
\hline
 18&$900\pm110$\\
\hline
 20&$790\pm100$\\
\hline
 25& $590\pm100$ \\
\hline
 27& $550\pm80$ \\
\hline
 29&$490\pm80$\\
\hline
 35&$290\pm60$ \\
\hline
\end{tabular}
\end{center}
\end{table} 
\begin{figure}[!h] 
\begin{center}
\includegraphics*[scale=0.44]{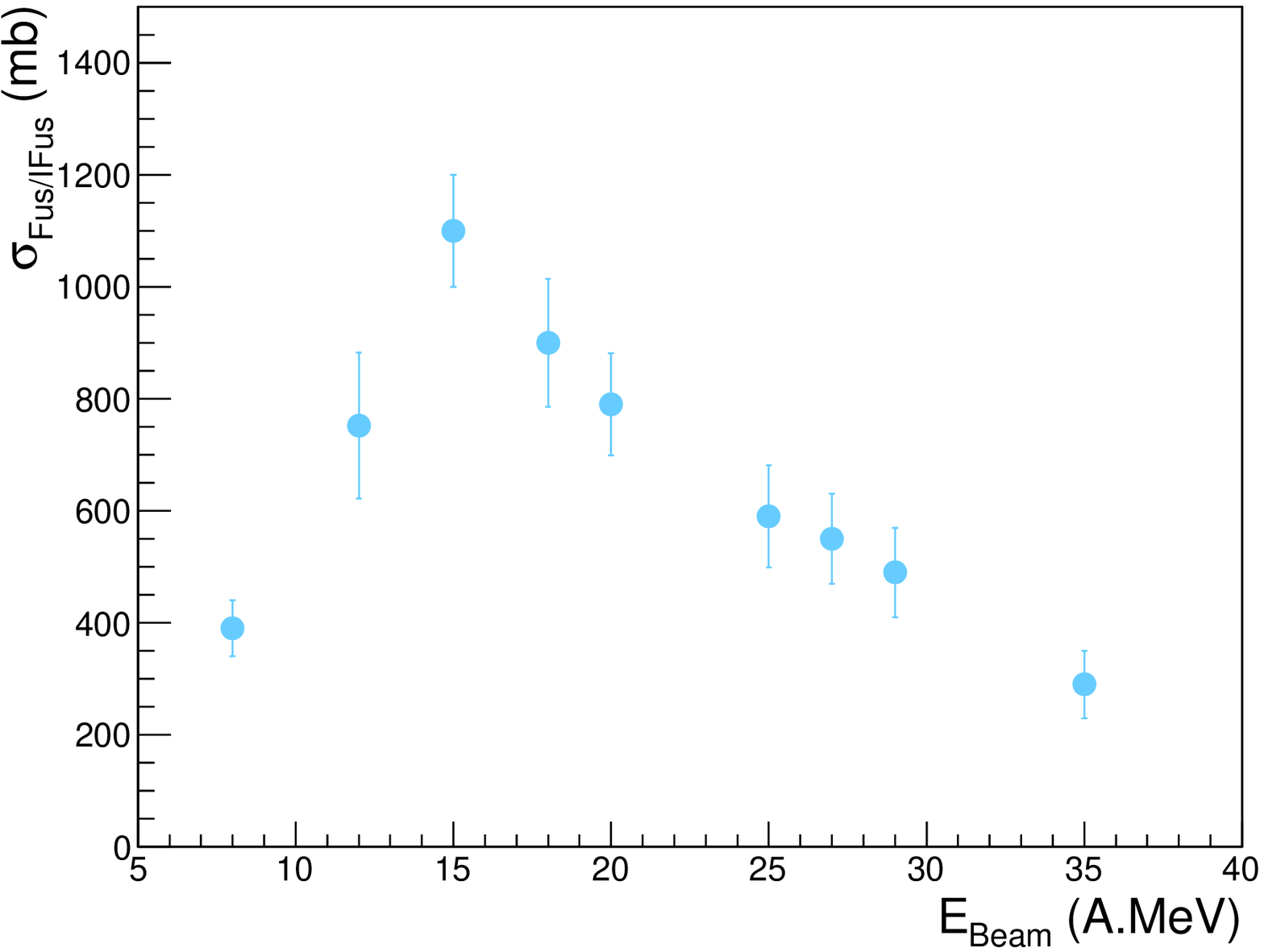}
\includegraphics*[scale=0.44]{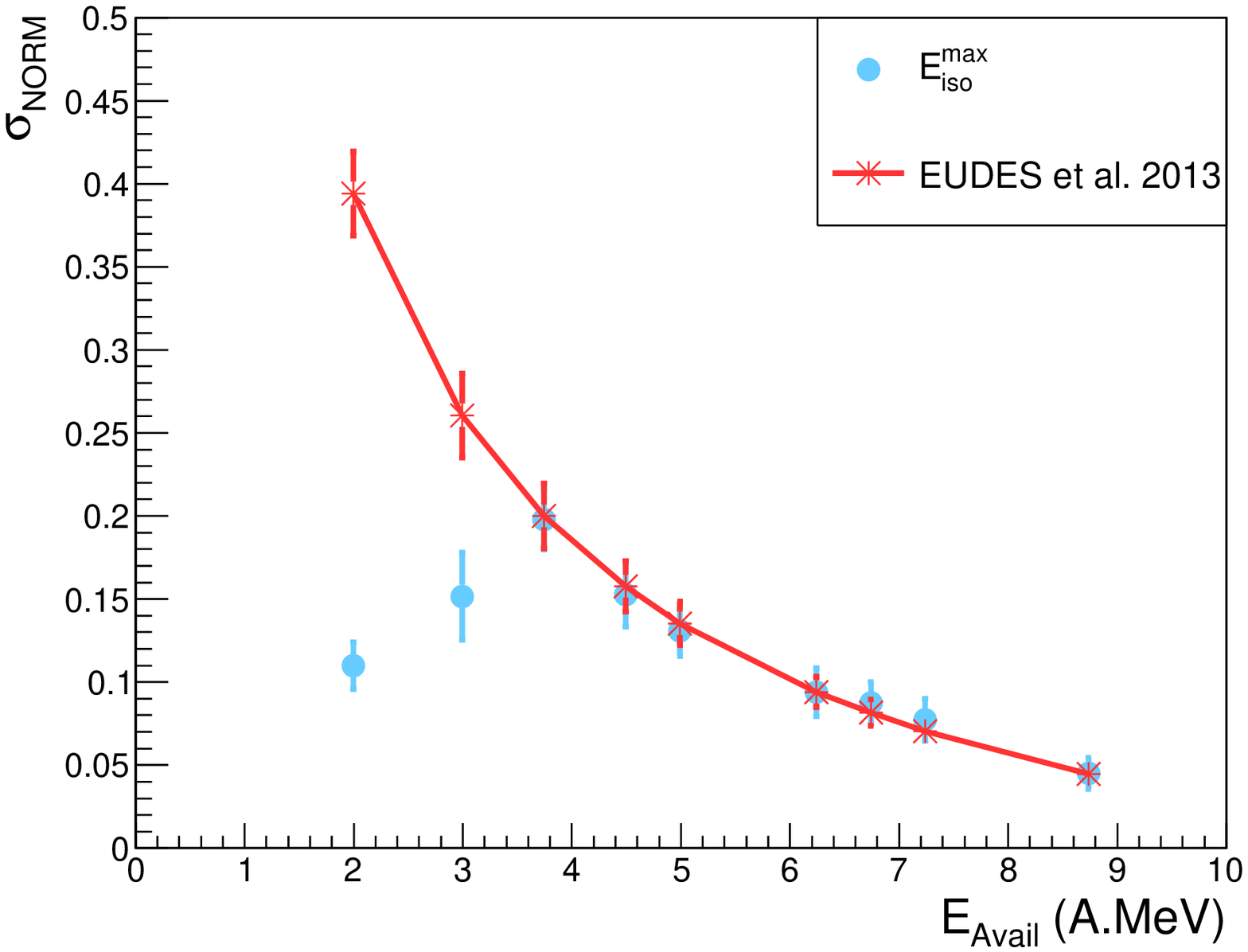}
\caption{Upper panel : experimental fusion cross section. Lower panel : fusion cross
section normalized for the Shen total reaction cross section.}
\label{FUScross}
\end{center}
\end{figure}

The fusion cross sections computed with the selections above exposed are shown in table \ref{fusion}. 
The error bars were evaluated by computing the cross section value corresponding to different selections on the heaviest charge
for higher beam energies. For lower energies it was useful to study the set composed by events without any fragment and compute the
cross section values with and without this set.\\
No correction for efficiency has still been applied : this could influence the cross section
results especially for low bombarding energies, were the compound formation may travel directly in the very forward
direction without being detected. However, in the meanwhile, this effect should be more dramatic as the beam energy increases since all the
products are more focalised in the forward direction.\\                                                                            
The fusion cross section values of table \ref{fusion} are shown in figure \ref{FUScross}. 
In the upper panel these values show a maximum at 15 A.MeV. 
One expects that for the beam energies of 8 and 12 A.MeV the values of the cross
sections are higher. Even if for 8 A.MeV the fusion hindrance, above discussed, would diminish the fusion probability, the
detector acceptance is suspected to be responsible for the loss of fusion/incomplete fusion events since all the residues with a forward angle 
lower than $3^{\circ}$ were lost. For increasing energies the probability for fusion events decreases 
rapidly : at  higher energies the transparency effets of the nuclear matter dominate as it was observed in
 \cite{lopezo},\cite{lop2014}. Because of its high kinetic energy the incident nucleus cannot anylonger succeed in forming a compound 
 nucleus with the
target and both are broken in several fragments during the collision.  \\
The lower panel of figure \ref{FUScross} shows the cross section values normalized to the Shen \cite{shen} values
 of the total reaction cross section previously discussed in order to compare them to those obtained for our system as described in 
 the work accomplished by P. Eudes and
 collegues in the references \cite{nuovoeudes}  
 which displays a systematic study on a large body of fusion data in order to
deduce a universal behavior.    
 The different systems were organized depending on
their size taking into account the mass asymmetry and data were plotted in
function of the available energy defined as :
\begin{equation}
E_{avail} = \frac{E_{lab}}{A_{proj}}\frac{A_{proj}A_{target}}{(A_{proj}+A_{target})^2}
\end{equation}
The authors deduced an homographic function starting from the ratio 
of the fusion cross section to the reaction cross section. 
In the present work the red curve with star symbols in figure \ref{FUScross} is the homographic function calculated for the 
$^{129}Xe+^{nat}Sn$ system using the parameters given in reference \cite{nuovoeudes}. \\
As one can see from figure \ref{FUScross}, the cross sections values are in fairly good agreement with the function from 
Eudes {\it{et al.}}. Although the homographic function in reference \cite{nuovoeudes} was deduced for light to intermediate systems and then extrapolated 
to heavy systems it seems to work fairly well also for this heavy system. On the red curve one can see the expected values for 8
and 12 A.MeV. These two lowest bombarding energies are the closest to the barrier and it would be important
to understand if their cross section are hindered because of the lower incident energy. 
A first preliminary insight came recently \cite{tesi}. A first simplified simulation based on the two step model \cite{shen1,shen2} applied to 
the system at 8 and 12 A.MeV beam energies (for $J=0$) gave fusion cross sections in agreement with the experimental ones above reported. The
question is still studied in order to take in account correctly the angular momentum.

%

\section{Discussion and Conclusions}

In this work the energy dependence of the experimental reaction cross section was displayed
and compared to different parameterizations. Data were found in fair agreement with those which take into account the effects due to
the neutron skin and to the Pauli blocking \cite{shen}.\par

Then we turned to the fusion cross section evaluation for which candidates for fusion events were selected with the help of the $E_{iso,max}$ 
observable. It is the first time
that such a study is accomplished, for a wide energy range from 8 to 35 A.MeV, on the quasi-symmetric heavy system
$^{129}Xe + ^{nat}Sn$.\par
The excitation function obtained shows a maximum around 15 A.MeV and then falls to lower values at increasing energies. Values
of the same order can be found in literature for mass intermediate systems and for lower bombarding energies \cite{tam} or
light systems at comparable energies \cite{lau} and references there in.\par 
For heavy colliding nuclei the compound sytem formed decays by fission which becomes the favorite exit channel
depending on the fissility of the compound system $Z^2/A$. Since in the diabatic hindrance model starting from a fusibility parameter $x_m =0.75$
\cite{ber,bolley} fusion becomes a dynamically hindered process, for $^{129}Xe + ^{nat}Sn$ with $x_m\simeq 9$, fission and
quasi-fission are clearly in competition. \par
As we discussed and showed, our method of computing the fusion cross section succeeds in excluding the DIC component. It
is clear that for this system there is a quasi-fission component. As already discussed, this process is slower than the 
DIC and does not proceed through the compound nucleus formation.  
To quantify it one should determine the characteristic decay times from 
angular distributions or solve isotopically the detected fragments \cite{toke}. This was beyond our purpose which was merely to supply a fusion upper bound for this system
in particular, scaling with the beam energies. However in a future FAZIA \cite{sal} could be able to detect the fragments with a good isotopic
resolution and so could be able to disentangle the different fusion-like mechanisms.  \\ 
The fusion cross sections of table \ref{fusion} were compared to a theoretical curve expected 
 to give an universal behavior. For all the energies but the lower ones (8 and 12 A.MeV) it was found a nice agreement with the
 universal behavior found by Eudes et al. \cite{nuovoeudes}. \\ 
 More data on quasi-symmetric and heavy systems would help to support the results of the present study. \\
The selection with the $E_{iso,max}$ observable revealed powerful in separating the different contributions for
central, semiperipheral and peripheral events. In particular the sample of semiperipheral collisions show few
characteristics reminding Deep Inelastic Collisions as it was displayed in the text. \par
As already quoted, the evident discrepancy from the general trend for lower energies (8 and 12 A.MeV) could be ascribed to the
intrinsic difficulty (the second inner fusion barrier needing an extra-push energy to be overcame) for heavy elements to form a true
compound nucleus. The acceptance of INDRA at very low angles (lower than $3^{\circ}$) complicates the 
analysis because of the loss of those
residues ejected in the very forward direction and only slightly deviated by the light particle evaporation process. 
To better understand and clarify this point simulations with a Montecarlo code are needed. In particular, simulations  
with the code HIPSE \cite{Dany} and GEMINI \cite{gem,john1} are currently in progress in order to better understand 
both the role of INDRA acceptance and of fusion hindrance. They will constitute the subject of a forthcoming article.



\begin{thebibliography}{0}
\bibitem{udo}W. U. Schr$\ddot o$drer {\it{et al.}}, Phys. Rep. 45, 167 (1978).
\bibitem{glassel}P.Gl$\ddot a$ssel {\it{et al.}}, Z.Phys.A - Atoms and Nuclei 310,189-216 (1983).
\bibitem{stefanini}A.A. Stefanini {\it{et al.}}, Z.Phys.A 351, 167-186 (1995).
\bibitem{charity}R.J. Charity {\it{et al.}}, Z.Phys.A - Hadrons and Nuclei 341, 53-73 (1991).
\bibitem{wilczynski}J. Wilczy\`nski {\it{et al.}},Phys. Rev. C 81067604 (2010).
\bibitem{bass0}R.Bass, Nucl. Phys. A 231, 45 (1974).
\bibitem{NGO}C.Ng\^o {\it{et al.}}, Nucl. Phys. A240 (1975) 353-364.
\bibitem{tam} B. Tamain, C. NG\^O, J. Peter and F. Hanappe, Nucl. Phys. A252 (1975) 187-207.
\bibitem{mosel}U. Mosel : Treatise on Heavy-Ion Science, Vol.2; Bromley (New
York Plenum, 1984)
\bibitem{lefort}M.Lefort and Ch. Ng\^o, Ann.Phys., (1978), V.3, 1.
\bibitem {udoB} Damped Nuclear Reactions, W.U.Schroeder and J.R.Huizenga,
Treatise on Heavy-Ion Science, Vol. , Bromley Ed., N.Y. (1984).
\bibitem{wilcz}J. Wilczy\`nski  Phys. Lett. B 47 124 (1973).
\bibitem{wilczb}J. Wilczy\`nski  Phys. Lett. B 47 484 (1973).
\bibitem{toke}J.Toke {\it{et al.}}, Nucl. Phys. A440 (1985) 327-365.
\bibitem{heush}B.Heusch {\it{et al.}}, Z. Phys. A 288 (1878) 391.
\bibitem{bernard} B. Borderie, M. Berlanger {\it{et al.}}, Z.Phys. A 299 (1981) 263.
\bibitem{back}B.B. Back, H. Esbensen, C.L. Jiang and K.E. Rehm, Rev.Mod.Phy., 86, (2014).
\bibitem{pete}J. P\'eter,C. Ng\^o,F. Plasil, M. Berlanger, and F. Hanappe, Nucl. Phys. A279 (1977) 110. 
\bibitem{itkisa} M.G. Itkis {\it{et al.}}, 2001 Proc. Int. Conf. on Fusion Dynamics at the extremes (Dubna) ed Yu Ts Oganessian and V I Zagrebaev
(Singapore : World Scientific) p. 93
\bibitem{itkis} M.G. Itkis {\it{et al.}}, Nucl. Phys. A 734,(2004) 136.
\bibitem{zag} V. Zagrebaev and W. Greiner, J. Phys. G : Nucl. Part. Phys. 31 (2005) 825-844.
\bibitem{Sahm} C.-C. Sahm {\it{et al.}}, Z.Phys. A 319, (1984) 113.
\bibitem{keller} J.G. Keller {\it{et al.}}, Nucl. Phys. A452,(1986) 173.
\bibitem{morawek} W. Morawek {\it{et al.}}, Z.Phys. A 341, (1991) 75.
\bibitem{Quint}A.B. Quint {\it{et al.}}, Z.Phys. A 346, (1993) 119.
\bibitem{swia1} W.J. Swiatecki, Phys. Scr. 24 (1981) 113-22.
\bibitem{swia2} W.J. Swiatecki, Nucl. Phys. A 376 (1982) 275-91.
\bibitem{bloki} J.P. Blocki {\it{et al.}}, Nucl. Phys. A 459 (1986) 145-72.
\bibitem{luk} A. Lukasiak, W. Cassing and W. Noremberg, Nucl. Phys. A426 (1984) 181-204.
\bibitem{ber} D. Berdichevsky {\it{et al.}}, Nucl. Phys. A502 (1989) 395c-404c.
\bibitem{Swia2003} W.J. Swiatecki, Acta Phys. Pol. B 34, (2003) 2049. 





\bibitem{pouthas1}
J.Pouthas {\it{et al.}} Nucl. Instr. and Meth. A357 (1995) 418.
\bibitem{pouthas2}
J.Pouthas {\it{et al.}} Nucl. Instr. and Meth. A369 (1996) 222.
\bibitem{parlog1}
M. P\^arlog {\it{et al.}} Nucl. Instr. and Meth. A482 (2002) 674.
\bibitem{parlog2}
M. P\^arlog {\it{et al.}} Nucl. Instr. and Meth. A482 (2002) 693. 
\bibitem{lukasik} J.Lukasik {\it{et al.}}, Phys. Rev. C 55 (1997) 1906-1916.
\bibitem{nat1} N.Marie {\it{et al.}}, Phys. Lett. B 391 (1997) 15-21.
\bibitem{nat2}N.Marie {\it{et al.}}, Phys. Rev. C 58 (1998), 256-269.
\bibitem{Neb} R.Nebauer {\it{et al.}}, Nucl. Phys. A 658 (1999) 67-93.
\bibitem{gou}D.Gourio {\it{et al.}}, Eur. Phys. J. A 7, 245-253 (2000).
\bibitem{ger}M.Germain {\it{et al.}}, Phys. Lett. B 488 (2000) 211-217.
\bibitem{boc}F.Bocage {\it{et al.}}, Nucl. Phys. A 676 (2000) 391-408.
\bibitem{meti}V.Metivier {\it{et al.}}, Nucl. Phys. A 672 (2000) 357-375.
\bibitem{bor}B.Borderie {\it{et al.}}, Phys. Rev. Lett. vol. 86 (2001) 3252-3255.
\bibitem{john1}J.D.Frankland {\it{et al.}}, Nucl. Phys. A 689 (2001) 940-964.
\bibitem{steck}J.C. Steckmeyer {\it{et al.}}, Nucl. Phys. A 686 (2001) 537-567.
\bibitem{belle}N.Bellaize {\it{et al.}}, Nucl. Phys. A 709 (2002) 367-391.
\bibitem{cussol}D.Cussol {\it{et al.}}, Phys. Rev. C 65 (2002) 044604.
\bibitem{hudan1}S. Hudan {\it{et al.}}, Phys. Rev. C 67 (2003), 064613.
\bibitem{colin}J.Colin {\it{et al.}}, Phys. Rev. C 67 (2003), 064603.
\bibitem{fev}A.Le Fevre {\it{et al.}}, Nucl. Phys. A 735 (2004) 219-247.
\bibitem{pia}S.Piantelli {\it{et al.}}, Phys. Lett. B 627 (2005) 18-25.
\bibitem{john2}J.D.Frankland {\it{et al.}},Phys. Rev. C 71 (2005) 034607.
\bibitem{pinch}M. Pinchon {\it{et al.}}, Nucl. Phys. A 779 (2006) 267.
\bibitem{lau}P.Lautesse {\it{et al.}}, Eur. Phys. J., A 27 (2006) 349-357.
\bibitem{tab}G.Tabacaru {\it{et al.}}, Nucl. Phys. A 764 (2006) 371-386.
\bibitem{nicol2007}N.Le Neindre {\it{et al.}}, Nucl. Phys. A 795 (2007) 47-69.
\bibitem{pia1}S.Piantelli {\it{et al.}}, Nucl. Phys. A 809 (2008) 11-128.
\bibitem{fev1}A.Le Fevre {\it{et al.}}, Phys. Lett. B 659 (2008) 807-812.
\bibitem{bonnet}E.Bonnet {\it{et al.}}, Nucl. Phys. A 816 (2009) 1-18.
\bibitem{lopezo} G. Lehaut {\it{et al.}}, Phys. Rev. Let. 104, 232701 (2010).
\bibitem{bonnet1}E.Bonnet et al., Phys. Rev. Lett. 105,(2010)14701.
\bibitem{bor1}B.Borderie {\it{et al.}}, Phys. Lett. B 723 (2013) 140-144.
\bibitem{gru}D.Gruyer {\it{et al.}}, Phys. Rev. Let. 110, 172701 (2013);
\bibitem{lop2014} O. Lopez {\it{et al.}}, Phys. Rev. C 90, 064602(2014).
\bibitem{ade}G.Ademard {\it{et al.}}, Eur. Phys. J., A (2014) 50:33.
\bibitem{gru1}D.Gruyer {\it{et al.}}, Phys. Rev. C 92, 064606(2015).

\bibitem {qu} R. Schiwietz  Nucl. Instr. and Meth. B 175-177 (2001) 125.
\bibitem{Bass} R. Bass, Nuclear Reactions with Heavy Ions, Springer-Verlag
Berlin Heidelberg New York 1980.
\bibitem{kox}S. Kox et al., Nucl. Phys. A420 (1984) 162; Nucl. Phys. A 420 (1984) 162; Phys. Rev. C 35 (1987) 1678.
\bibitem{TRI} R.K. Tripathi, F.E. Cucinotta and J.W. Wilson, Nasa Technical Paper 3621 (1997).
\bibitem{shen} W.Q. Shen {\it{et al.}}, Nucl. Phys. A491 (1989) 130-146.
\bibitem{pippo}G. Ademard {\it{et al.}}, EPJA 50, 33 (2014).
\bibitem{remi}R.Bougault et al., in preparation (2015).
\bibitem{phair}L. Phair {\it{et al.}}, Nucl. Phys. A 548 (1992) 489.
\bibitem{pla} E. Plagnol {\it{et al.}}, Phys. Rev. C 61, 014606 (1999).
\bibitem{john}Private communication, December 2015.
\bibitem{bot}A. Botvina {\it{et al.}}, Nucl. Phys. A475 (1987) 663.
\bibitem{DiToro}M. Di Toro, A. Olmi and R. Roy, Eur. Phys. J. A 30, 65-70 (2006).
\bibitem{flot}M.Mjahed, thesis LPC Clermont-Ferrand, (1987).
\bibitem{viola} V.E.Viola,Jr., Nucl. Data Sect. A1, 391 (1966).
\bibitem{nuovoeudes}P. Eudes et al., EPL, 104 (2013) 22001.
\bibitem{tesi}Hongliang Lu, Doctoral Thesis of the University of Caen (France) (2015)
\bibitem{shen1}C.Shen, G. Kosenko and Y.Abe, Phys. Rev. C 66 (2002) 061602.
\bibitem{shen2}C.Shen {\it{et al.}}, Int. J. Mod. Phys. E 17,(2008) 66. 
\bibitem{bolley} Shen C.W., Abe Y., Li ,Q. F., Boilley D., Sci. China Ser G, 2009, 52 (10) 1458-1463;
Shen C. W. {\it{et al.}}, Phys. Rev. C 83, 054620 (2011).
\bibitem{sal}F.Salomon {\it{et al.}}, Jou. Ins. Vol. 11 (2016).
\bibitem{Dany}D. Lacroix {\it{et al.}}, Phys. Rev. C 69, 054604 (2004). 
D.Lacroix and D. Durand, arXiv:nucl-th/0505053; D.Lacroix {\it{et al.}},Phys. Rev. C 71, 024601 (2005).
\bibitem{gem}R.Charity {\it{et al.}}, Nucl. Phys. A476 (1988) 516-544.
\bibitem{john1}J.Frankland IWM 2016






%


%

\end{thebibliography}
\end{document}